\documentclass[twocolumn,showpacs,preprintnumbers,amsmath,amssymb]{revtex4}

\usepackage{graphicx}
\usepackage{dcolumn}
\usepackage{bm}

\begin{document}
\title{Circular magneto-optical trap for neutral atoms}
\author{Makoto Morinaga}
\email{morinaga@ils.uec.ac.jp}
\affiliation{Institute for Laser Science,
University of Electro-Communications and CREST,
Chofu-shi, Tokyo, 182-8585 Japan}

\date{\today}

\begin{abstract}
We propose and experimentally demonstrate
a novel scheme to magneto-optically trap neutral atoms in
a ring shaped trap that can be used to transfer atoms into a
circular magnetic trap with high density. This inturn enables to
evaporatively cool atoms and study the behaviour of ultra cold gases in
a periodic
2-dimensional potential. The circular magneto-optical trap itself is
also of
interest to investigate the properties of magneto-optical trap of
deformed shape, such as reduction of photon-reabsorption.
\end{abstract}


\maketitle

\section{Introduction}
Ultra cold gases tightly confined in a 2-dimensional (2-D) potential
are expected to exhibit various features which are not seen in a
bulk 3-D system,
and are extensively investigated both theoretically and experimentally.
These features include anomalous quantum fluctuations of
quasi-condensate
\cite{Ho}, the fermionic
behaviour of a bosonic gas in the Tonks-Girardeau regime
\cite{Kinoshita},
or the quantized conductance through such a confining potential
\cite{Prentiss}.
In most experimental studies linear magnetic traps (MT) or optical traps
of finite length are used
and small changes of the potential in the longitudinal direction
are unavoidable.
A ring MT on the other hand
has in principle a flat potential
along the trap ring, and because it is a periodic system,
additional phenomena such as persistent currents or Josephson
effects are expected to be observed.
For the practical side,
the study of evaporative cooling process in 2-D potential is
of importance to achieve a continuous atom laser \cite{Dalibard}.
Several groups have so far realized magnetic storage rings
\cite{Chapman}-\cite{Stamper-Kurn}
that are mainly aimed to construct a Mach-Zehnder type interferometer
for the high precision gyroscope application
and thus the atomic density should be kept low to avoid the unwanted
atom-atom interaction.
In fact, these traps are loaded from a conventional
3-D magneto-optical trap (MOT) or 3-D MT and
the density of atoms is insufficient to start evaporative
cooling when atoms are uniformly spread in the trap.
In this letter we propose a new type of MOT in which
atoms are trapped along a circle.
Because of the large trapping volume and the trap shape matching,
an efficient loading of circular MT from this ring MOT should be
possible, which will enable the evaporative cooling of atoms
in the circular MT.
We also expect that the problem of fragmentation of the condensate
due to the irregularities of the trap potential will be reduced by
rotating atoms in the circular trap, that is not possible for
the linear (box-like in the longitudinal direction) 1-D potential.
\\
As an MOT of reduced dimension,
suppression of photon-reabsorption and high density trap
are also expected \cite{Castin}.

\section{Basic idea}
\begin{figure*}[!htbp]
\resizebox{15cm}{!}{%
\includegraphics{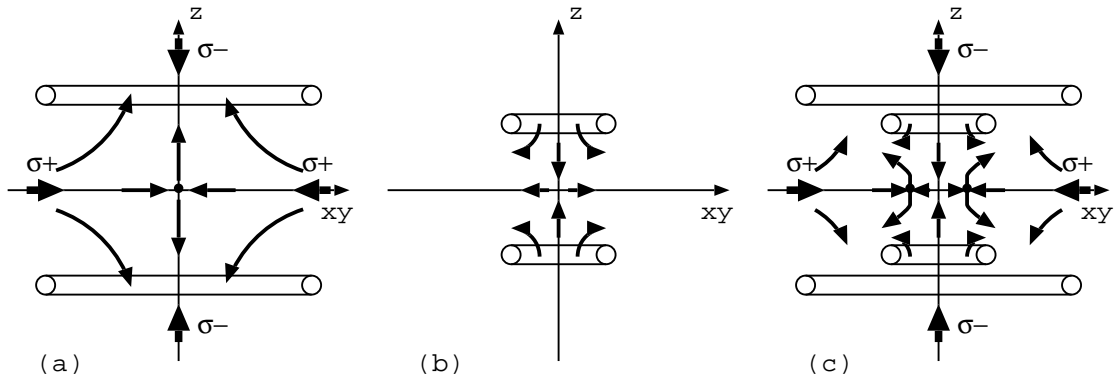}
}
\caption
{Configuration of the magnetic field and cooling beams of MOTs.
(a) normal MOT. (b) small inner quadrupole coils pair to modify the
magnetic field. (c) ring MOT.
}
\label{MOTs}
\end{figure*}

A ring MOT is realized by modifying the magnetic field of a
conventional 3-D MOT. Fig. \ref{MOTs}(a) is a schematic drawing
of the popular 6-beam MOT.
An anti-Helmholtz coils pair generates a quadrupole magnetic
field at the center and
the trapping laser beams approaching the center of the trap
parallel (anti-parallel) to the magnetic field
have $\sigma_+$ ($\sigma_-$) polarization so that the atoms are
trapped at the zero of the magnetic field.
Now we add a small inner quadrupole coils pair (fig. \ref{MOTs}(b))
to this system.
By putting appropriate currents to this coils pair in the
opposite direction, the original zero of the magnetic field
is expelled from the center to form a ring.
With the trapping laser beams unchanged, atoms are now pulled
toward this ring (fig. \ref{MOTs}(c)).
Because the atoms are trapped at the zero of the magnetic field,
we can switch to the ring MT just by turning off the
laser beams.
In MT, spin-flip of trapped atoms can be avoided
by using TORT (time-orbiting ring trap) technique proposed in
\cite{Arnold}, or by putting a current carrying wire on the
symmetry axis ($z$-axis) which generates an azimuthal magnetic
field $B_\theta$.
When the trap with this current carrying wire is operated as an MOT,
$B_\theta$ causes imbalance in the scattering
force along the trap ring and atoms begin to rotate.
The final rotation speed (mean velocity
$v_{rot}$) is determined by the balance between the Zeeman shift
by $B_\theta$ and the Doppler shift by $v_{rot}$.
This can be used to generate a laser cooled atomic ensemble that has
finite mean velocity, which might be useful for some applications,
such as the observation of the glory scattering \cite{glory}.

\section{Some mathematics}
\begin{figure*}
\resizebox{15cm}{!}{%
\includegraphics{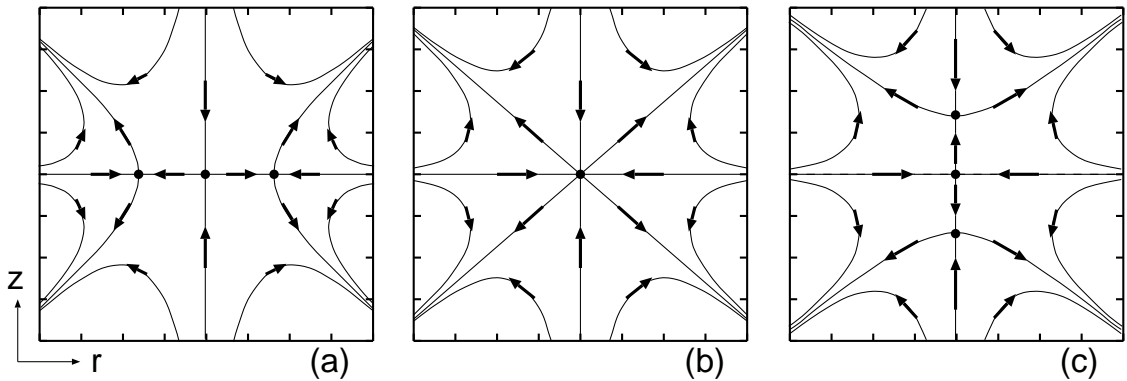}
}
\caption{Profile of the magnetic field.
Magnetic field lines are shown for (a) $a_2a_4>0$ (ring MOT),
(b) $a_2=0$ (octapole field), and (c) $a_2a_4<0$ (``double MOT'').}
\label{profile}
\end{figure*}

In vacuum (i.e. in the absence of currents)
a static magnetic field $\vec B$ is derived from
a (magnetic) scalar potential $\phi$: $\vec B=-\vec\nabla\phi$.
Let us consider the magnetic field generated by a set of
coils which has
rotation symmetry around $z$-axis and
anti-symmetry for $z$-axis inversion. Thus
$\phi=\phi(z,r)$,
$\phi(-z,r)=\phi(z,r)$ with $r\equiv\sqrt{x^2+y^2}$.
Now we expand $\phi$ near the origin (center of the trap)
to the 5th order in $z$ and $r$:
\begin{equation}
\phi=a_2\phi_2+a_4\phi_4
\label{s_potential}
\end{equation}
with $\phi_2=z^2-\frac 12r^2,\ \phi_4=z^4-3r^2z^2+\frac 38r^4$
\footnote{Generally, $n$th order term is given by
$\phi_n=\sum_{\nu =0}^{[\frac n2]}(-1)^\nu
\left(\frac 1{\nu !}\right)^2\frac{n!}{(n-2\nu)!}\ z^{n-2\nu}
\left(\frac r2\right)^{2\nu}$ \cite{focusing}.
}.
From this, we calculate
\begin{equation}
B_r(0,r)=-\frac 32a_4r(r^2-\frac{2a_2}{3a_4}),
\label{eq_br}
\end{equation}
\begin{equation}
B_z(z,0)=-4a_4z(z^2+\frac{a_2}{2a_4}).
\label{eq_bz}
\end{equation}
If $a_2a_4>0$, from (\ref{eq_br}), we see that there is
a circle of zero magnetic field of radius
$r_{trap}=\sqrt{\frac{2a_2}{3a_4}}$.
Field gradient on this ring is calculated as
$\partial_rB_r|_{(0,r_{trap})}=-\partial_zB_z|_{(0,r_{trap})}
=-2a_2$.
From (\ref{eq_bz}) on the other hand, in case
$a_2a_4<0$, there will be two additional
points of zero magnetic field on $z$-axis at
$z_d=\pm\sqrt{-\frac{a_2}{2a_4}}$.
Again field gradients on these points are calculated as
$\partial_rB_r|_{(z_d,0)}=-\frac 12\partial_zB_z|_{(z_d,0)}
=-2a_2$. Because the field gradients are the same
for both points,
we can simultaneously trap atoms on these points
(``double MOT'') by appropriately choosing the direction
of currents, or the helicity of the trapping laser beams.
\\
The profile of the magnetic field is depicted in fig. \ref{profile}
for $a_2a_4>0$, $a_2=0$
(octapole field\footnote{It is known that a 3-D
MOT with $2n$-pole magnetic field is possible only for $n=2$
\cite{thesis}.}),
and $a_2a_4<0$
\footnote{In a cylindrically symmetric system, field lines
can be drawn as contour lines of the function
$f(z,r)=\int_0^r \rho\partial_z\phi(z,\rho)\ d\rho$.
}.
\\
In a more general case where the the system has no anti-symmetry
under the inversion of $z$-axis, odd order terms in $z$ also come in
the equation (\ref{s_potential}). 
This will rotate the principal axes of the quadratic field in $zr$-plane, and
if rotated by $\pi/4$ (which is the case when the system is symmetric under
$z\rightarrow -z$), the restoring force toward the ring disappears.

\section{Experiment}
\begin{figure}
\resizebox{7cm}{!}{%
\includegraphics{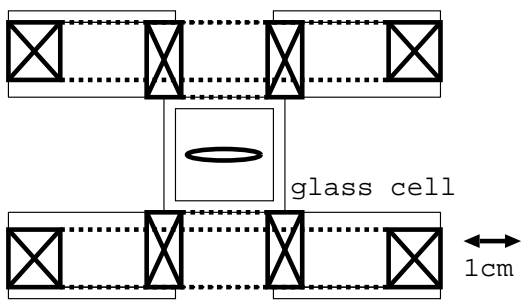}
}
\caption{Design of the trap coils. A trap glass cell is sandwiched
between pairs of outer and inner coils. The coils are cooled by water.}
\label{coils}
\end{figure}

\begin{figure*}
\resizebox{6cm}{!}{%
\includegraphics{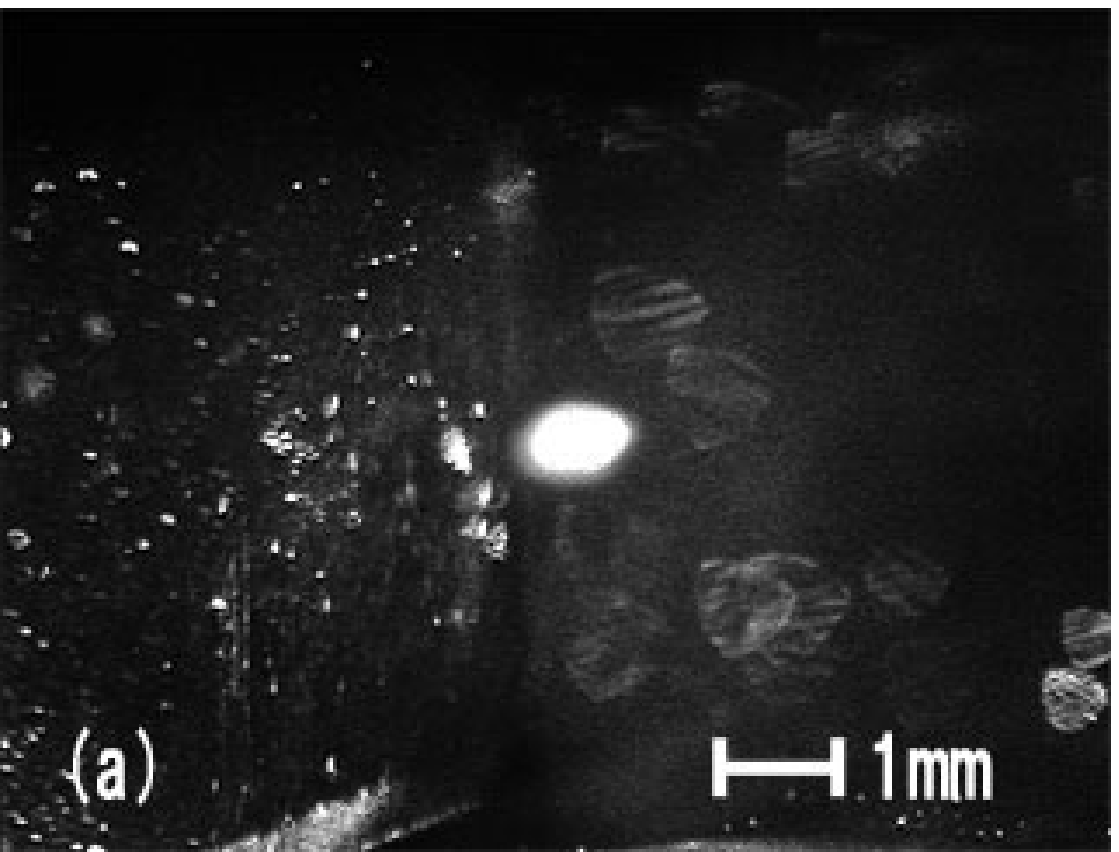}
}
\hspace{10mm}
\resizebox{6cm}{!}{%
\includegraphics{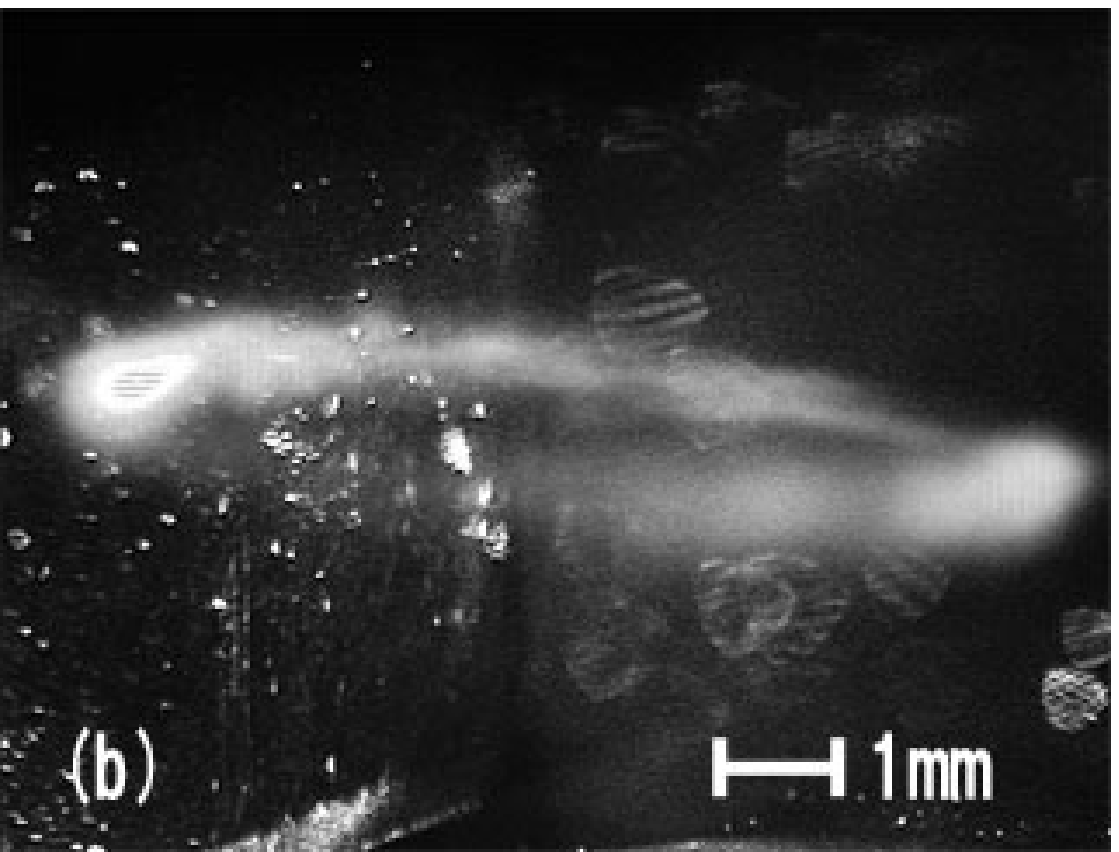}
}
\caption{(a) Normal MOT ($I_L=9$A, $I_S=0$A,
and $\partial_zB_z(0,0)=-2\partial_rB_r(0,0)=18$ Gauss/cm).
(b) Circular MOT ($I_L=18$ A, $I_S=$6 A,
and $\partial_zB_z(0,r_{ring})=-\partial_rB_r(0,r_{ring})=4.5$ Gauss/cm
with $r_{ring}=3.8$ mm).}
\label{exp}
\end{figure*}

\begin{figure*}
\resizebox{15cm}{!}{%
\includegraphics{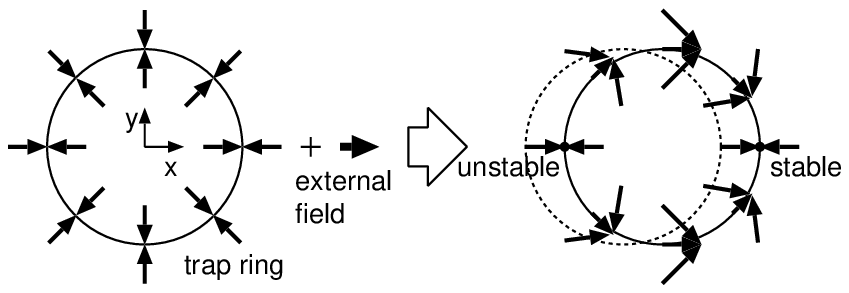}
}
\caption{A slice of magnetic field in $xy$-plane under the perturbation
by a constant external field in $+x$-direction.}
\label{slice}
\end{figure*}

We have performed a preliminary experiment to prove the principle
of this trap.
A sodium dispencer (SAES Getters) is placed 15cm away from the center of the
trap, and atoms are catched directly by the MOT without using Zeeman slower.
We use a ring dye laser (Coherent CR699-21 with Rhodamine 6G) for the
trapping laser (about 50mW in each arms with diameter $\sim$15mm) and an
electro-optical modulator to generate a 1.77GHz sideband for the
repumping.
The design of the trapping coils is shown in fig. \ref{coils}.
Parameters of these coils in (\ref{s_potential}) are given by
\begin{eqnarray}
a_2=A_{out2} I_{out} - A_{in2} I_{in}
\nonumber
\\
a_4=A_{out4} I_{out} - A_{in4} I_{in} \nonumber
\end{eqnarray}
with
$A_{in2}=6.5$ Gauss cm$^{-1}$A$^{-1}$,
$A_{out2}=2.0$ Gauss cm$^{-1}$A$^{-1}$,
$A_{in4}=1.7$ Gauss cm$^{-3}$A$^{-1}$,
$A_{out4}=0.0$ Gauss cm$^{-3}$A$^{-1}$
($I_{out}$ and $I_{in}$ are the currents flowing in the
outer and inner coils, respectively).
Maximum currents for the both coils are $I_{max}=18$A.
In fig. \ref{exp} we show pictures (fluorescence images) of MOTs under
the normal and the circular MOT conditions.
The inhomogeneity of the atomic cloud density
of the circular MOT can be explained
by the insufficient beam quality of the trapping
lasers. Another possible account for the inhomogeneity
comes from the fact that the circle
of zero magnetic field can easily be destroyed by an external stray
magnetic field (or by the small misaligment of the trap coils). 
For example, consider the perturbation by a small uniform magnetic field
in $+x$-direction (a slice of the magnetic field in $xy$-plane is shown in
fig. \ref{slice}). On $x$-axis, the points of zero magnetic field are
just shifted by the external field in $+x$-direction. In the other area,
however, there are no local zero points and points of local minima
form again a circle, along which atoms are accumulated toward the
point of zero magnetic field on the right handside of the circle.
We also note in fig. \ref{exp}(b) that the axis of the trap ring
is tilted slightly.
This is due to the mismatch of the centers of the vertical
trapping beams, which was not a serious concern for the
conventional MOT.

\section{Conclusion and outlook}
We have proposed and experimentaly demonstrated a novel method
to magneto-optically trap neutral atoms in a circular trap
that can be used to load laser cooled atoms into a circular
magnetic trap. This method opens up a path to generate and
investigate 1-dimensional cold gas with periodic boundary condition.
We are now working on constructing a new setup using electromagnets to
achieve much tighter confinement.

\begin{acknowledgments}
We thank C. Maeda for assistance with the experiment, and V. I. Balykin
and T. Kishimoto for useful discussions.
This work is partly supported by the 21st Century COE program of the
University of Electro-Communications on ``Coherent Optical Science''
supported by the Ministry of Education, Culture, Sports, Science
and Technology.
\end{acknowledgments}

\end{document}